\documentclass[gdf]{sifrev}        
\usepackage{geometry}
\usepackage{amsmath}           
\usepackage{graphicx}          
\usepackage[italian]{babel}
\usepackage{bm}
\usepackage{hyperref}

\begin{document}

\title{La forma della Terra: una lezione sulla gravit\`a Newtoniana}


\author{Marco Ruggieri\inst{1} and Guang Xiong Peng\inst{1}\inst{2}}          
 
\institute{College of Physics, University of Chinese Academy of Sciences, Yuquanlu 19A, Beijing 100049, China
\and
Theoretical Physics Center for Science Facilities, Institute of High Energy Physics, Beijing 100049, China}      

\branch{W}      

\vol{42}                                  
\issue{2}				  

\month{Month year}                        


\maketitle

\section{Introduzione}     

Uno degli aspetti fondamentali della didattica della Fisica, in particolare dei corsi di Fisica generale
sia nelle scuole secondarie che universitari, \`e quello di sottolineare come il metodo scientifico sia lo strumento
pi\`u importante nell'affermazione delle teorie scientifiche. Ovvero, l'esposizione di nozioni
pi\`u o meno astratte \`e sempre affiancata alla verifica sperimentale delle leggi pi\`u importanti,
e si invitano gli studenti ad apprendere non solo la formulazione teorica delle leggi ma anche
il modo in cui queste sono verificate sperimentalmente.
Questo mette in luce la complessa, ma allo stesso tempo rigorosa, natura di una teoria scientifica,
che alla formulazione matematica di una legge affianca la misura sperimentale
che verifichi la bont\`a della suddetta legge. 

Uno dei metodi che riteniamo efficace nella presentazione di una teoria scientifica \`e quello di
porre lo studente di fronte ad un problema concreto, affrontando il quale lo studente stesso apprenda
il modo di lavorare di uno scienziato e sviluppi uno spirito critico che gli permetta poi di risolvere problemi pi\`u
complicati. 
Quello riportato in questo articolo \`e un esempio
che, con una esposizione riproducibile in una classe di Fisica generale
sia di scuola secondaria che di universit\`a,
evidenzia  come affrontare un problema attraverso il metodo scientifico, fissando un modello teorico e
focalizzando l'attenzione sulla scelta delle quantit\`a  fisicamente rilevanti per il problema, da cui
si intuisce l'utilit\`a nel calcolarle, fare il calcolo utilizzando una strategia
appropriata (numerica o analitica) e confrontare poi la predizione teorica del modello con 
l'esperienza. 

In particolare ci proponiamo in questa lezione di mostrare come,
assumendo che sia valida la teoria della gravitazione di Newton,
questa sia {\em incompatibile} con una forma non sferica della Terra analizzando del moto dei gravi in caduta libera.
Nonostante la forma sferica (o a geoide) della Terra sia ben conosciuta 
e non deve essere messa in discussione in una lezione di Fisica,
riteniamo interessante mostrare come ipotizzando la correttezza della gravit\`a Newtoniana
e l'applicazione della stessa
permettano di comprendere gli effetti che si dovrebbero misurare nel momento in cui
la Terra non fosse sferica.  Come spiegheremo pi\`u in dettaglio nel seguito,
la semplice analisi del moto dei gravi nel campo gravitazionale di una Terra non sferica
mostra come tale moto sia in netto contrasto con quello osservato sperimentalmente.
Assunta quindi la gravit\`a Newtoniana come teoria della gravit\`a, 
l'analisi del moto mostra l'incompatibilit\`a tra questa ed una forma
non sferica del pianeta. 

In una lezione di Fisica generale il problema che presentiamo qui andrebbe impostato nel modo seguente: 
dopo aver spiegato
agli studenti la formulazione Newtoniana della gravitazione, e aver messo in luce come questa
sia una teoria scientifica corredata sia di una descrizione matematica dell'interazione gravitazionale
che di accurate verifiche sperimentali, nonch\`e come questa permetta di fare previsioni quantitative
su grandezze fisiche misurabili sperimentalmente, si pu\`o suscitare negli studenti la curiosit\`a su che tipo
di grandezza fisica si potrebbe studiare per evidenziare la compatibilit\`a della gravit\`a
Newtoniana con la forma sferica della Terra.
L'individuazione di un'opportuna grandezza fisica porta poi al calcolo della stessa usando vari modelli,
ognuno caratterizzato da una certa geometria per la distribuzione di massa della Terra.
Infine, il confronto dei risultati ottenuti con quelli empirici permette di escludere uno o pi\`u modelli,
lasciando in effetti come unica possibilit\`a compatibile con la gravit\`a Newtoniana quella di una Terra sferica.
La quantit\`a fisica che proponiamo di studiare in questa lezione \`e l'accelerazione di gravit\`a
sulla superficie della Terra. Mostreremo infatti che questa grandezza, oltre ad essere misurabile sperimentalmente
ed accessibile qualitativamente tramite l'esperienza quotidiana,
sia sensibile alla distribuzione geometrica della massa che la produce.
In particolare, l'accelerazione di gravit\`a \`e utile perch\`e anche l'aspetto qualitativo
del moto di un grave \`e fortemente influenzato dalla forma del pianeta. 

La presentazione in classe del tema qui riportato pu\`o inoltre essere supportata da una 
lezione applicativa di laboratorio di calcolo numerico. Infatti, durante l'esposizione degli argomenti
qui discussi sar\`a necessario fare riferimento ad alcuni integrali di volume per il campo
gravitazionale sulla superficie della Terra, che sono facilmente calcolabili utilizzando tecniche 
Monte Carlo che certamente non fanno parte del programma di un corso di Fisica generale
ma che possono facilmente essere parte di un corso di calcolo numerico o di un laboratorio di informatica.  
Nella presentazione in classe dunque si pu\`o illustrare direttamente il risultato del calcolo
del campo gravitazionale, dopo aver mostrato agli studenti che tipo di integrale \`e coinvolto nell'analisi
del problema, rimandare ai corsi di calcolo o di laboratorio la scrittura di un codice numerico per 
la sua concreta valutazione.

\section{Un richiamo sulla gravit\`a di Newton}

In questa sezione richiamiamo molto brevemente la formulazione dell'interazione
gravitazionale Newtoniana.
Per una trattazione pi\`u approfondita si veda ad esempio~\cite{panareo,MS,DS}.

Consideriamo due masse puntiformi che chiamiamo $m$ ed $M$ 
poste a distanza $r$ nello spazio; la forza gravitazionale che la massa $M$ esercita
su $m$ \`e data da
\begin{equation}
\bm{F}= -G \frac{m M}{r^2} \frac{\bm r}{r},
\label{eq:vecF}
\end{equation}
dove $\bm r$ denota il vettore che congiunge le due particelle e $G$ denota la costante di gravitazione universale;
ovviamente la forza che esercita $1$ su $2$
sar\`a uguale in modulo ed opposta in verso. Notiamo che $\bm F$ \`e diretta sempre in direzione opposta al vettore
$\bm r$: la forza \`e dunque sempre attrattiva.
La costante $G$ che compare nella  (\ref{eq:vecF}) \`e detta {\it costante di gravitazione universale};
il suo valore numerico \`e stato misurato indirettamente per la prima volta da Cavendish nel 
1798 \cite{cavendish}
ed il valore oggi accettato dalla comunit\`a scientifica internazionale \`e (tralasciando
per semplicit\`a il valore dell'incertezza)
\begin{equation}
G = 6.67\times 10^{-11}~\mathrm{\frac{N\times m^2}{kg^2}}.
\label{eq:Gnew}
\end{equation}

Notiamo che definendo il campo gravitazionale 
generato dalla massa $M$, e denotandolo con $\bm g$, come
\begin{equation}
\bm g = -G \frac{M}{r^2} \frac{\bm r}{r},
\label{eq:cg1}
\end{equation}
allora la forza di interazione gravitazionale (\ref{eq:vecF}) pu\`o essere riscritta nella forma
\begin{equation}
\bm F = m \bm g;
\label{eq:vecF2}
\end{equation}
il campo $\bm g$ \`e anche chiamato, in questo contesto, accelerazione di gravit\`a in quanto rappresenta l'accelerazione
che un corpo sperimenterebbe qualora fosse soggetto alla forza  (\ref{eq:vecF2}).
Assumendo $M \gg m$ gli effetti di $m$ sul moto di $M$ sono trascurabili. 
Diciamo allora che il corpo di massa $M$ \`e la sorgente del campo gravitazionale
mentre $m$ \`e una particella test che sonda il campo prodotto da $M$ senza perturbarlo.

Nel caso in cui la sorgente del campo gravitazionale 
sia estesa nello spazio \`e necessario generalizzare l'equazione (\ref{eq:vecF}).  
Tale generalizzazione \`e immediata in quanto nella formulazione di Newton della gravit\`a, il campo gravitazionale
gode del principio di sovrapposizione: per calcolare la forza di gravit\`a prodotta da un insieme di 
$N$ masse puntiformi $\{M_i\}$ su una massa che chiamiamo massa test e denotiamo con $m$, 
\`e sufficiente sommare  l'equazione (\ref{eq:vecF}) su tutte le masse in $\{M_i\}$, per cui 
\begin{equation}
\bm F(\bm r_P) = -G m \sum_{i=1}^N \frac{M_i}{|\bm r_P - \bm r_i|^2}
\frac{\bm r_P - \bm r_i}{|\bm r_P - \bm r_i|},
\label{eq:33}
\end{equation}
dove $\bm r_P$ corrisponde al vettore posizione della massa $m$ ed $\bm r_i$ al vettore posizione
della massa $i-$esima appartenente ad $\{M_i\}$. Nel caso in cui il set $\{M_i\}$ sia sostituito da una distribuzione
continua di massa caratterizzata da una densit\`a $\rho(\bm r)$ l'equazione (\ref{eq:33}) si generalizza come
\begin{equation}
\bm F(\bm r_P) = -G m \int_\Omega\! d^3r \frac{\rho(\bm r)}{|\bm r_P - \bm r|^2}
\frac{\bm r_P - \bm r}{|\bm r_P - \bm r|},
\label{eq:33a}
\end{equation}
dove $\Omega$ rappresenta la regione di spazio occupata dalla distribuzione di massa, e la densit\`a \`e definita
come
\begin{equation}
M = \int_\Omega\! d^3r \rho(\bm r),
\end{equation}
essendo $M$ la massa totale del sistema. 

Notiamo che analogamente alla (\ref{eq:vecF2}), 
grazie alla (\ref{eq:33a}) la forza di gravit\`a sulla massa puntiforme $m$ pu\`o essere scritta nella forma
\begin{equation}
\bm F(\bm r_P) = m \bm{g}(\bm r_P),
\label{eq:33c}
\end{equation}
dove
\begin{equation}
\bm g(\bm r_P) = -G \int_\Omega\! d^3r \frac{\rho(\bm r)}{|\bm r_P - \bm r|^2}
\frac{\bm r_P - \bm r}{|\bm r_P - \bm r|},
\label{eq:33d}
\end{equation}
\`e l'accelerazione di gravit\`a che si misurerebbe in $\bm r_P$
e la (\ref{eq:33c}) assume la forma della forza peso standard.
Nel seguito calcoleremo $\bm g(\bm r_P)$ essendo ovvio che nota questa quantit\`a la forza  
$\bm F(\bm r_P)$ si ottiene banalmente dalla (\ref{eq:33c}).

\section{Come la geometria influenza la gravit\`a}
In questa sezione introduciamo tre modelli per la geometria della Terra, che chiameremo
rispettivamente Terra sferica, Terra a disco (o piatta) e Terra a calotta.
Per ognuno di questi tre modelli ci proponiamo di calcolare la forza di gravit\`a
sulla superficie esercitata su un grave di massa $m$: questo equivale a calcolare
l'accelerazione di gravit\`a. 
Lo scopo di questo calcolo \`e quello di mostrare come l'accelerazione di gravit\`a
dipende, sia qualitativamente che quantitativamente, dalla geometria assunta per il pianeta.

\subsection{La gravit\`a sulla Terra sferica}

Il primo modello discutiamo \`e quello di una Terra sferica.
Come \`e noto la Terra \`e  un geoide, 
ovvero un solido tridimensionale ellissoidale dalla superficie irregolare:
nonostante il raggio equatoriale medio sia maggiore di quello polare medio,
la variazione fra i due  \`e di appena dello $0.3\%$;
il raggio medio terrestre \`e $R_T\approx 6370$ km.
Inoltre, le irregolarit\`a della superficie della Terra dovute ad esempio alle montagne sono irrilevanti
per il problema: basta considerare che le pi\`u alte montagne della Terra, ovvero quelle della catena
dell'Himalaya, hanno altezza rispetto al livello del mare inferiore ai 9 km, per cui le fluttuazioni della distribuzione di massa
rispetto sulla superficie di una sfera con raggio uguale ad $R_T$
sono dell'ordine dello $0.1\%$. 
Per i nostri scopi \`e sufficiente dunque considerare una Terra
sferica con raggio pari al raggio medio terrestre. 
Per calcolare la densit\`a usiamo anche la massa $M_T\approx 5.97\times 10^{24}$ kg; 
in questo caso la densit\`a \`e data da
$
\rho_T = M_T/V_T \approx 5.5\times 10^3~\mathrm{kg/m^3}, 
$
dove $V_T = 4\pi R_T^3/3$ denota il volume della Terra sferica. 
Come vedremo, e come dovrebbe essere fatto notare allo studente, queste assunzioni per quanto
forti sono sufficienti a riprodurre il valore medio dell'accelerazione di gravit\`a misurata sperimentalmente,
$g_{\mathrm{fen}}\approx 9.81$ m/sec$^2$.

\begin{figure}[t!]
\begin{center}
\includegraphics[scale=0.35]{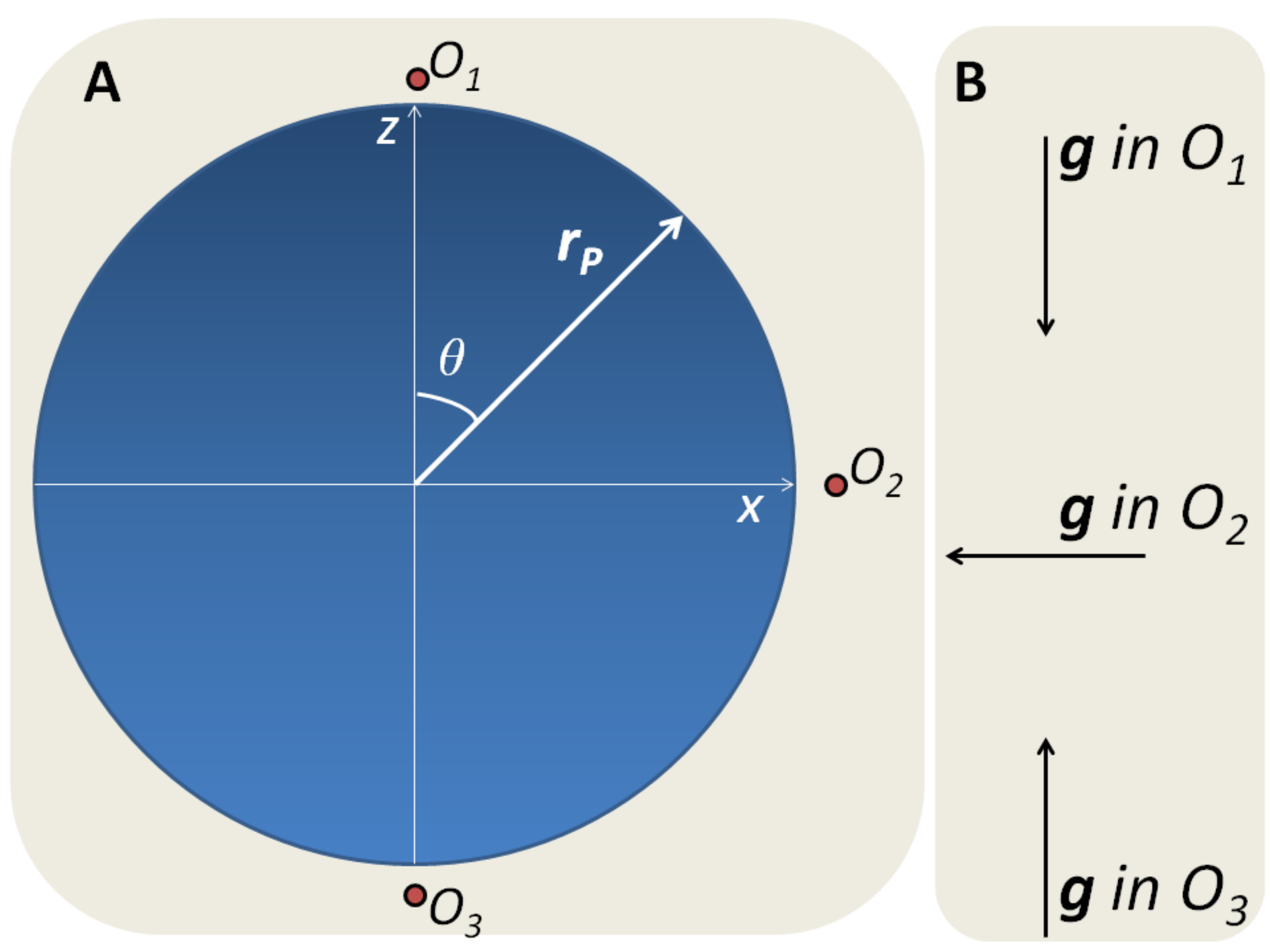}
\end{center}
\caption{\label{fig:1}Pannello A: Sezione trasversale della Terra e definizione del sistema di riferimento. Nella figura,
$O_1$, $O_2$ ed $O_3$ corrispondono ai tre osservatori menzionati nel testo. 
Pannello B: Accelerazione di gravit\`a misurata dagli osservatori.}
\end{figure}

Nel pannello A della Fig.~\ref{fig:1} \`e rappresentata una sezione longitudinale
della Terra sferica, ed il sistema di riferimento 
$x-z$ che useremo per il calcolo dell'accelerazione di gravit\`a con la definizione dell'angolo $\theta$.  
Abbiamo inoltre definito tre osservatori: $O_1$ per $\theta=0$ situato al Polo Nord,
$O_2$ per $\theta=\pi/2$ situato all'equatore ed $O_3$ per $\theta=\pi$ al Polo Sud.   
Lo scopo che ci proponiamo \`e quello di calcolare la forza di gravit\`a che un osservatore misurerebbe sulla
superficie della Terra tonda spostandosi con continuit\`a dalla posizione occupata da $O_1$ a quella occupata da $O_3$.

In generale, nota la distribuzione di massa definita tramite la densit\`a, 
l'equazione (\ref{eq:33a}) il calcolo della forza di gravit\`a esercitata sulla massa puntiforme $m$
nel punto $\bm r_P$ richiede il calcolo di un integrale di volume che, a meno di alcuni casi di geometrie semplici,
pu\`o essere calcolato solo numericamente. Nel caso della Terra sferica, assumendo una densit\`a indipendente
da $\bm r$, il teorema del flusso di Gauss permetterebbe il calcolo immediato del campo gravitazionale
sulla superficie terrestre; useremo qui per\`o direttamente
l'equazione (\ref{eq:33a}), come controllo della strategia numerica usata per il calcolo dell'integrale 
che nel caso di altri modelli di Terra \`e pi\`u complicato.

L'integrale nella (\ref{eq:33d}) si pu\`o calcolare numericamente usando un metodo Monte Carlo standard.
Per la massa $m$ possiamo scrivere
\begin{equation}
\bm{r}_P = (R_T\cos\phi\sin\theta,R_T\sin\phi\sin\theta,R_T\cos\theta),
\end{equation}
dove stiamo assumendo che $m$ si muova sulla superficie per cui il modulo del raggio vettore
coincide con $R_T$; inoltre possiamo supporre che $\phi=0$ per cui
\begin{equation}
\bm{r}_P = (R_T \sin\theta,0,R_T\cos\theta);
\end{equation}
si vede quindi che essendo $R_T$ fissato, l'unica variabile da cui dipende la forza \`e l'angolo $\theta$
definito nella figura \ref{fig:1}.

\begin{figure}[t!]
\begin{center}
\includegraphics[scale=0.35]{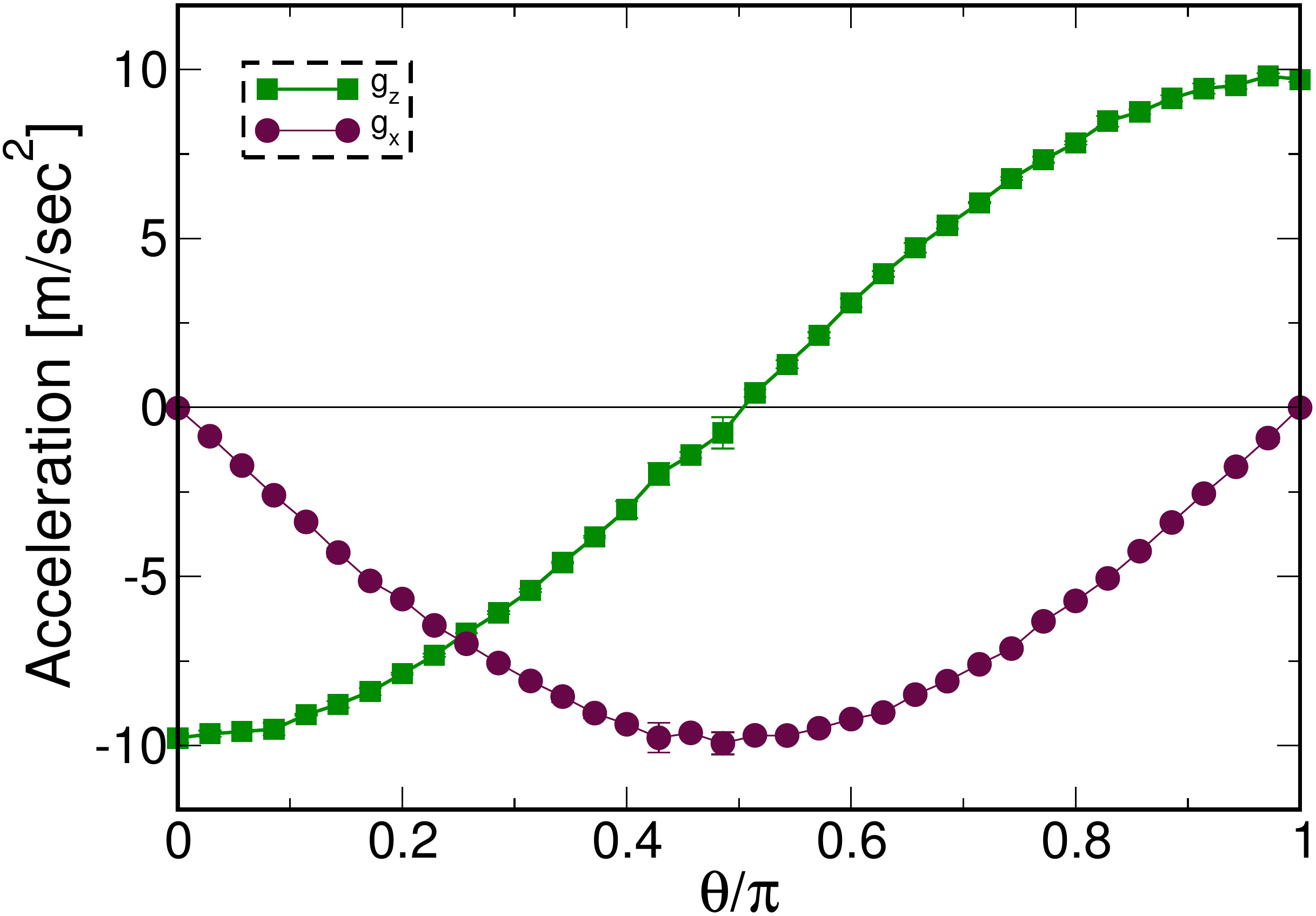}
\end{center}
\caption{\label{fig:2}Componenti dell'accelerazione gravitazionale in funzione dell'angolo $\theta$.}
\end{figure}

Nella Fig.~\ref{fig:2} mostriamo le componenti $g_z$ e $g_x$ dell'accelerazione di gravit\`a
nel caso della Terra sferica, in funzione dell'angolo $\theta$ espressa in radianti; 
le barre di errore, la cui larghezza \`e comunque comparabile con la dimensione dei punti riportati,
sono dovute alla natura probabilistica del calcolo dell'integrale con le tecniche Monte Carlo.

Osserviamo subito, come test dell'integrazione numerica  della (\ref{eq:33d}),
che il valore dell'accelerazione di gravit\`a calcolata \`e in ottimo accordo con quello fenomenologico, 
$g_{\mathrm{fen}}$.
Ricordiamo che stiamo 
considerando la Terra come una sfera perfetta, trascurando quindi lo schiacciamento ai poli
e la correzione dovuta alla forza centrifuga che introducono una dipendenza di $\bm g$
dalla latitudine. Tali correzioni non influenzano comunque in modo considerevole
il valore di $g$: la sua massima variazione con la latitudine infatti \`e dell'ordine dello $0.5\%$.
Confrontando $\bm g$ calcolata in $O_1$ e $O_2$ notiamo che 
le componenti $g_z$ e $g_x$ si scambiano ma l'accelerazione risultante \`e sempre diretta
verso il centro della Terra, ovvero ha solo componente ortogonale alla superficie stessa.

Osserviamo infine che per l'osservatore al Polo Sud, $O_3$,  
l'accelerazione di gravit\`a \`e ancora diretta verso il centro della Terra, anche se la componente cambia segno
in quanto nel sistema di riferimento in figura \ref{fig:1} la $\bm g$ al Polo Sud \`e concorde alla direzione
dell'asse $z$. Nel pannello B della \ref{fig:1} mostriamo i vettori $\bm g$ per i tre osservatori.
Si pu\`o quindi riassumere il risultato ottenuto scrivendo
\begin{equation}
\bm a_n =-\sqrt{g_z^2 + g_x^2}\bm n,~~~\bm a_t =0,
\label{eq:opo}
\end{equation}
dove $\bm a_n$ e $\bm a_t$ rappresentano le componenti normale e tangenziale dell'accelerazione di gravit\`a
rispettivamente, ed $\bm n$ denota il versore normale alla superficie.

\subsection{La gravit\`a sulla Terra a disco}

\begin{figure}[t!]
\begin{center}
\includegraphics[scale=0.35]{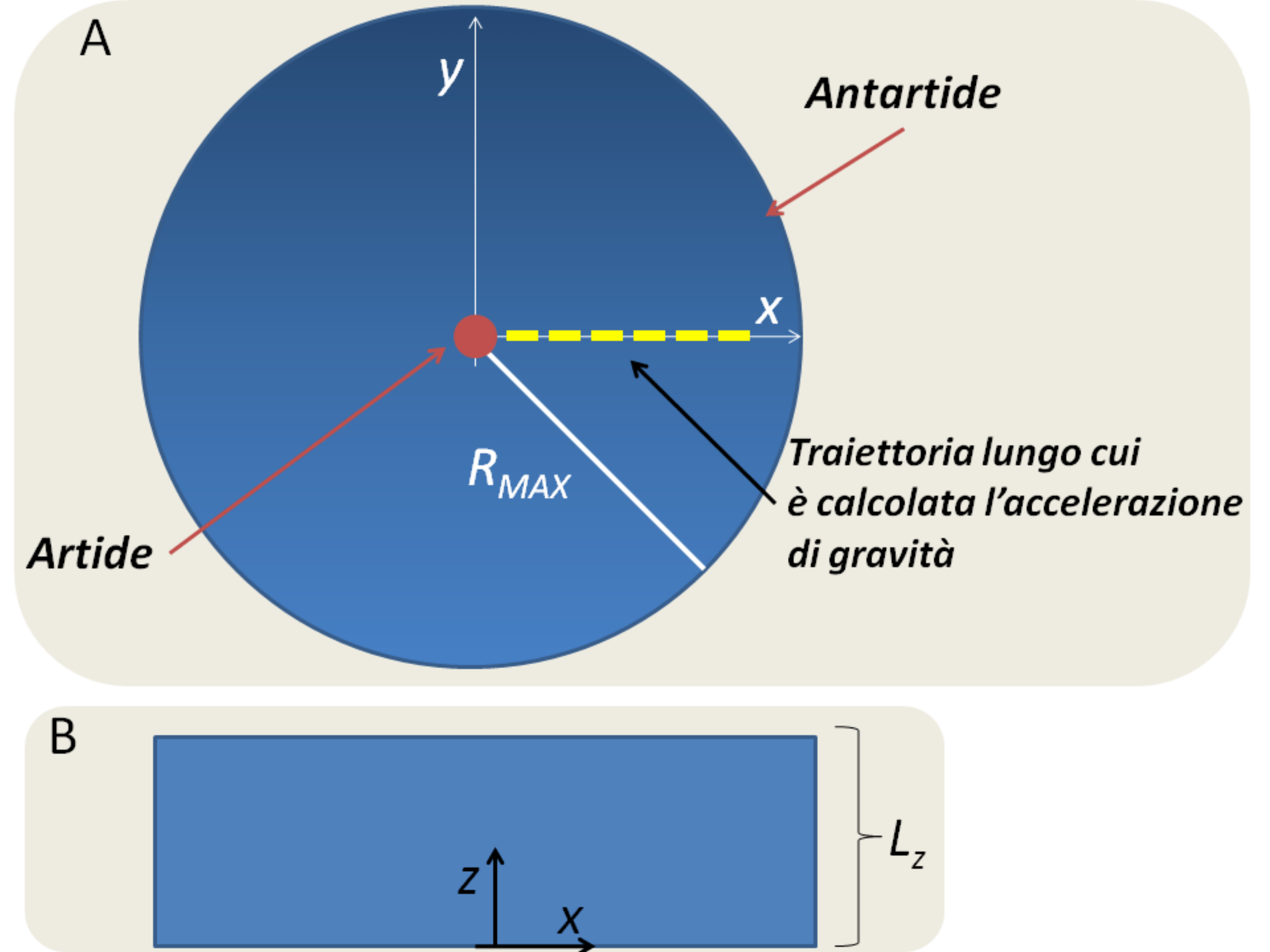}
\end{center}
\caption{\label{fig:1d} Modello di Terra a disco. I pannelli A e B rappresentano
rispettivamente la visione dall'alto e laterale. }
\end{figure}

Calcoliamo ora l'accelerazione di gravit\`a sulla superficie della Terra a disco.
Nella figura~\ref{fig:1d} mostriamo il modello usato per la presente analisi.
I pannelli A e B corrispondono alla vista dall'alto e laterale del modello rispettivamente. 
Sono inoltre evidenziate le due lunghezze fondamentali del modello, ovvero il raggio della Terra 
a disco $R_{\mathrm{MAX}}$ e lo spessore $L_z$.
In figura \`e anche indicata, per completezza, la traiettoria lungo la quale \`e calcolata l'accelerazione di gravit\`a,
rappresentata dalla linea gialla tratteggiata.
In questo modello di Terra, il Polo Nord corrisponderebbe al centro del cerchio mentre
i bordi corrisponderebbero all'Antartide. Chiaramente non vogliamo mettere assolutamente in discussione
la sfericit\`a della Terra e non vogliamo dare alcun peso scientifico al modello qui considerato:
ricordiamo che lo scopo della presente lezione \`e quello di stuzzicare lo spirito critico dello studente,
mostrandogli come una semplice considerazione sulle accelerazioni e sul moto dei gravi permette di
escludere il modello a disco (o piatto) della Terra.

Il calcolo della $\bm g$ che compare nelle (\ref{eq:33c}) e (\ref{eq:33d}) nel caso della Terra a disco
procede in maniera analoga a quello per la Terra sferica presentato nella sezione precedente.  
Assumiamo che la densit\`a della Terra a disco sia la stessa di quella della Terra sferica;
inoltre, $R_{\mathrm{MAX}}$ pu\`o essere facilmente fissato in quanto conosciamo la distanza
che deve essere percorsa per andare dal Polo Nord, situato al centro della Terra piatta, al bordo che 
rappresenterebbe l'Antartide, muovendosi lungo un meridiano: tale distanza
\`e pari a $\pi R_T$ dove $R_T$ rappresenta il raggio della Terra sferica. Infine, per fissare $L_z$
richiediamo che il modello dia un'accelerazione di gravit\`a al Polo Nord in accordo con il valore 
sperimentale $g_{\mathrm{fen}}$: troviamo
\begin{equation}
R_\mathrm{MAX}\approx2\times 10^4~\mathrm{km},~~~
L_z \approx 4.73\times 10^3~\mathrm{km}.
\label{eq:Lz_1}
\end{equation}

\begin{figure}[t!]
\begin{center}
\includegraphics[scale=0.35]{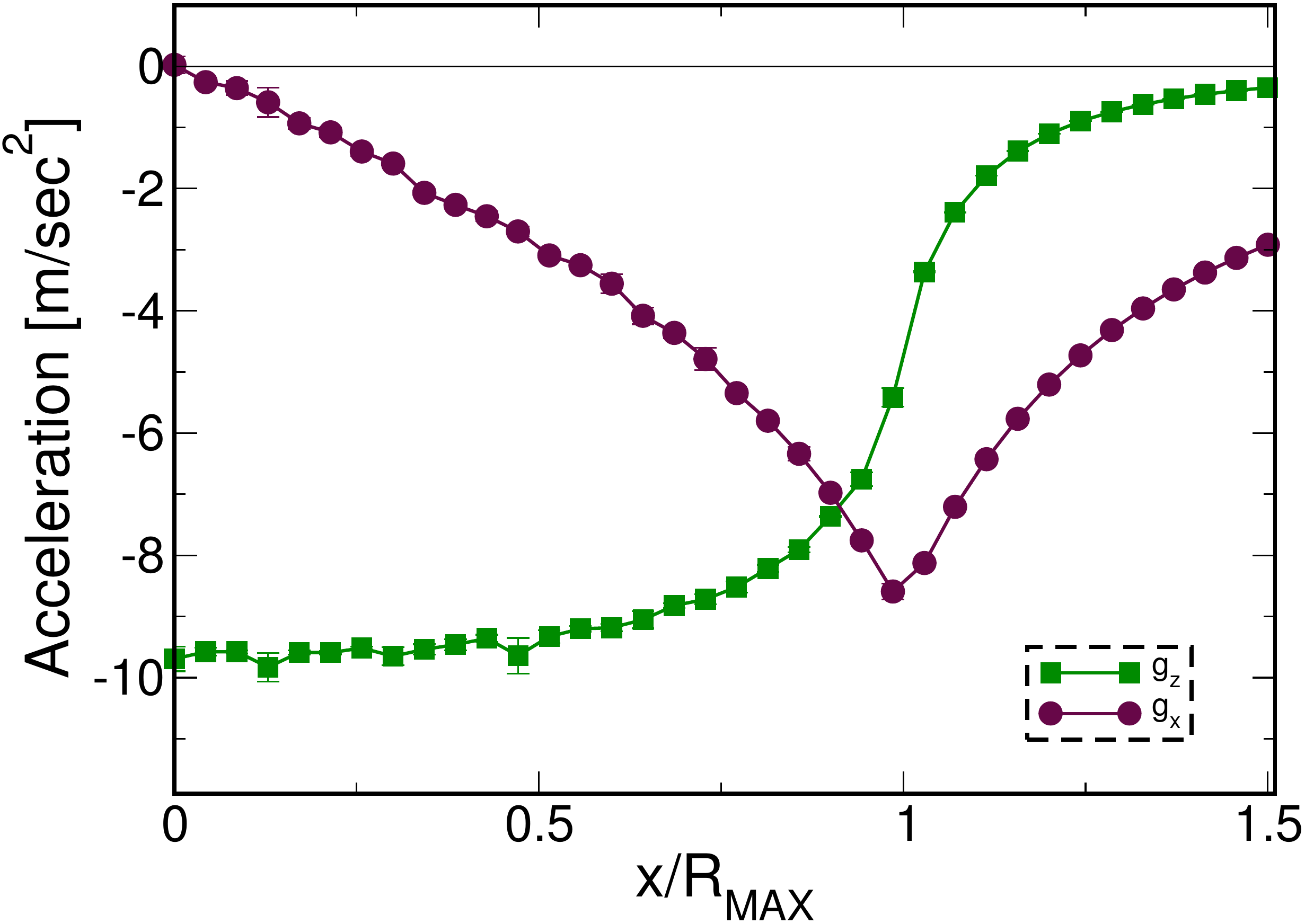}
\end{center}
\caption{\label{fig:4}Componenti normale, $g_z$, e tangenziale, $g_x$, dell'accelerazione di gravit\`a
sulla superficie della Terra a disco. Il sistema di coordinate \`e definito in figura~\ref{fig:1d}.}
\end{figure}

Nella figura~\ref{fig:4} mostriamo le componenti normale, $g_z$, e tangenziale, $g_x$, dell'accelerazione di gravit\`a
sulla superficie della Terra a disco. L'accelerazione \`e calcolata lungo la traiettoria indicata dalla linea gialla
nella figura~\ref{fig:1d}. 
In particolare $x=0$ corrisponde al Polo Nord, mentre $x=R_{\mathrm{MAX}}$ al bordo del disco.
Analogamente a quanto definito per la Terra sferica, 
vedi equazione (\ref{eq:opo}), abbiamo
\begin{equation}
\bm a_n = \bm g_z~,~~~\bm a_t=\bm{g}_x.
\label{eq:nt2}
\end{equation}
Osserviamo che
spostandosi sulla superficie lungo un meridiano dal centro della Terra a disco fino al suo ipotetico bordo, 
si sviluppa una componente tangenziale dell'accelerazione di gravit\`a, orientata verso il Polo Nord. 
Questo \`e evidente dai dati mostrati nella figura~\ref{fig:4}: la componente $g_x$ \`e nulla
al Polo Nord, che corrisponde ad $x=0$, ed \`e diversa da zero per $x>0$,
diventando inoltre paragonabile in intensit\`a alla componente $g_z$. 
Questo risultato \`e in netto contrasto con quello ottenuto per il
modello della Terra sferica. 
Deduciamo quindi che \`e possibile, attraverso una misura
di accelerazioni, discernere fra Terra a disco e Terra sferica. 
Infine, osserviamo che la componente tangenziale che si svilupperebbe ai bordi
della Terra a disco, essendo orientata verso il centro agirebbe come una forza di
richiamo tanto pi\`u intensa quanto pi\`u ci si avvicina ai bordi. Questo avrebbe un effetto
sia sulla stabilit\`a gravitazionale del pianeta, che sulla superficie libera degli oceani.
Senza entrare nel dettaglio di questi problemi, sottolineiamo solo che la presenza
della componente tangenziale dell'accelerazione di gravit\`a produrrebbe un moto di un
grave in caduta libera diverso da quello osservato sperimentalmente, come discuteremo
in una sezione successiva.

\begin{figure}[t!]
\begin{center}
\includegraphics[scale=0.35]{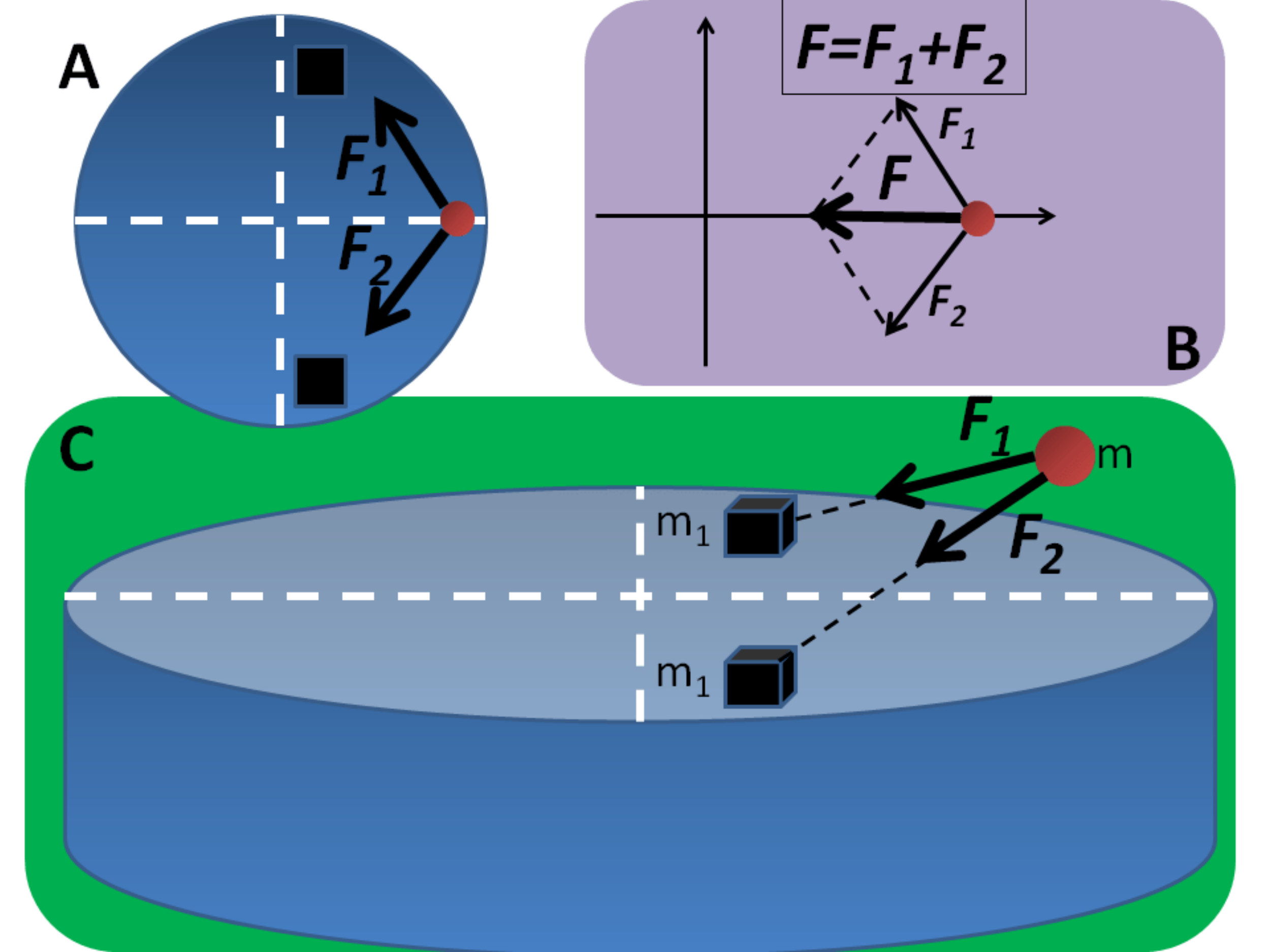}
\end{center}
\caption{\label{fig:qual}Composizione delle forze di attrazione gravitazionale sulla superficie
della Terra a disco. In particolare, i cubi neri rappresentano due elementi di volume di massa
$m_1$ che agiscono gravitazionalmente sul grave di massa $m$. 
Il pannello A mostra una proiezione del sistema sul piano della superficie, mentre il pannello C
mostra una visione in prospettiva. Infine, il pannello B mostra la composizione delle forze $F_1$ 
ed $F_2$ ed evidenzia l'esistenza di una componente tangenziale della forza risultante.}
\end{figure}

Qualitativamente, il risultato sulla forza di gravit\`a sulla superficie
della Terra a disco pu\`o essere compreso aiutandosi con la figura~\ref{fig:qual}. Nel pannello C
della figura mostriamo la Terra a disco in prospettiva: le masse $m_1$ corrispondono
a quelle di due elementi di volume diametralmente opposti che esercitano ognuno
l'attrazione gravitazionale, $\bm F_1$ ed $\bm F_2$,
nei confronti del grave di massa $m$. Nel pannello A della  figura~\ref{fig:qual}
mostriamo una proiezione del sistema sul piano della superficie della Terra. 
Nel pannello B mostriamo la composizione delle forze nel suddetto piano:
la forza risultante $\bm F = \bm F_1 + \bm F_2$ ha una componente non nulla
orientata verso il centro della superficie: questa componente \`e esattamente  
la componente tangenziale dell'attrazione che abbiamo discusso sopra.

\subsection{La gravit\`a sulla Terra a calotta}

\begin{figure}[t!]
\begin{center}
\includegraphics[scale=0.35]{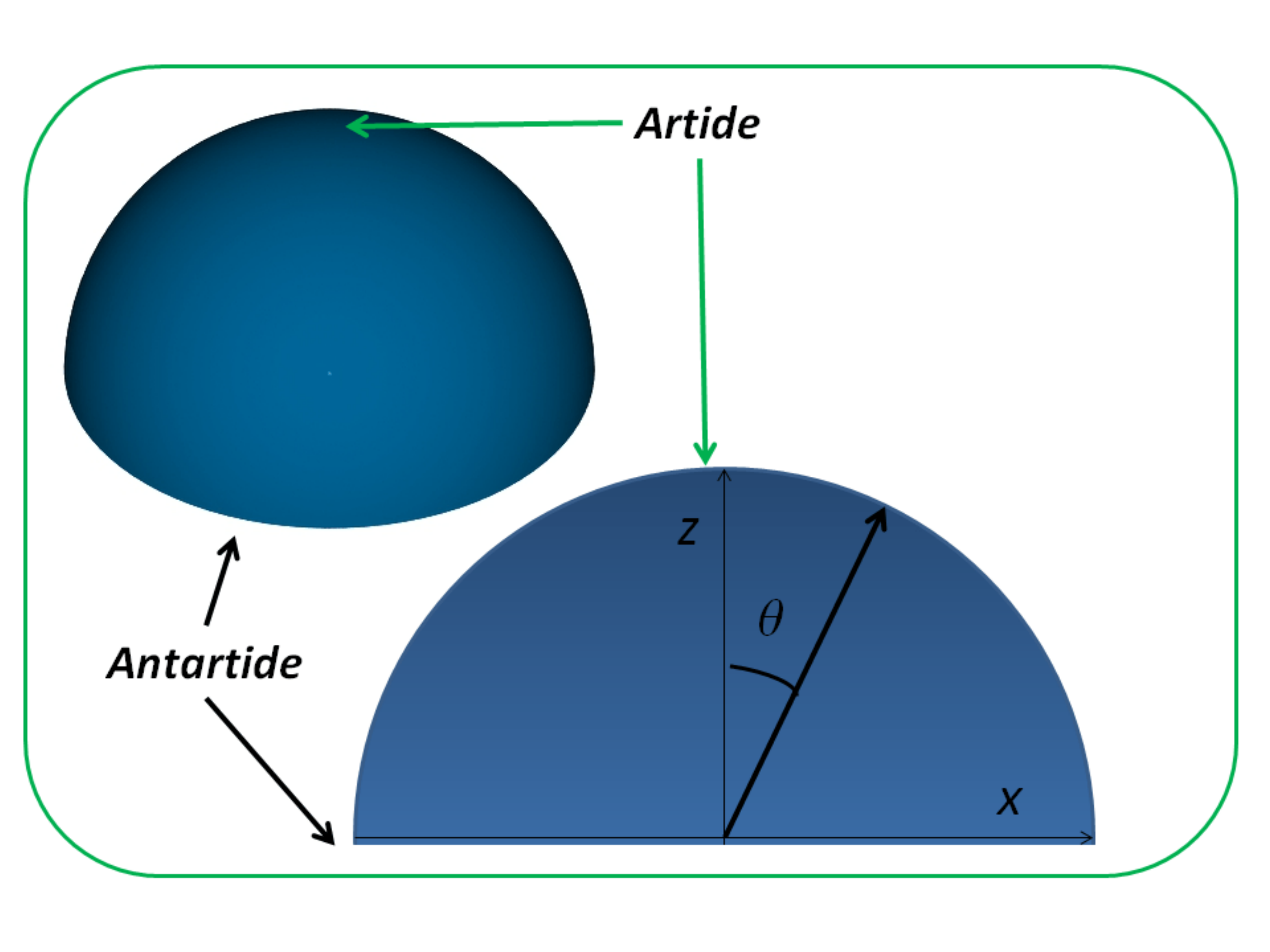}
\end{center}
\caption{\label{fig:7}Modello di Terra a calotta. }
\end{figure}

Consideriamo come ultimo modello quello di una Terra a forma di calotta sferica, 
si veda il pannello in alto a sinistra in figura~\ref{fig:7}. La forma geometrica che consideriamo qui
\`e quella di una semisfera perfetta; notiamo che aumentando il raggio di curvatura della semisfera \`e possibile passare con
continuit\`a da questa al caso della Terra a disco esaminato prima.
Geometricamente il problema \`e simile a quello del calcolo dell'accelerazione di gravit\`a per
il modello Terra tonda: anche in questo caso calcoliamo l'accelerazione di gravit\`a in funzione
dell'angolo $\theta$, vedi pannello in basso a destra in figura~\ref{fig:7}.
Data la geometria del modello per\`o l'angolo $\theta$ misurato in radianti
varia da $\theta=0$, che corrisponde al Polo Nord, a $\theta=\pi/2$ che invece corrisponderebbe
al bordo della semisfera occupato dall'Antartide.

Ancora una volta, per semplicit\`a assumiamo una densit\`a uniforme pari alla densit\`a media della Terra tonda,
ovvero $\rho_T\approx 5.5\times 10^3$ kg/m$^3$. Questa ipotesi lascia come unico parametro libero
del modello il raggio di curvatura  della semisfera, che denotiamo con $R_{\mathrm{MAX}}$ e che fissiamo 
richiedendo che l'accelerazione di gravit\`a per $\theta=0$
sia numericamente in accordo con il valore sperimentale
$g_\mathrm{fen}$: questo porta a $R_{\mathrm{MAX}}\approx 8030$ km.

\begin{figure}[t!]
\begin{center}
\includegraphics[scale=0.35]{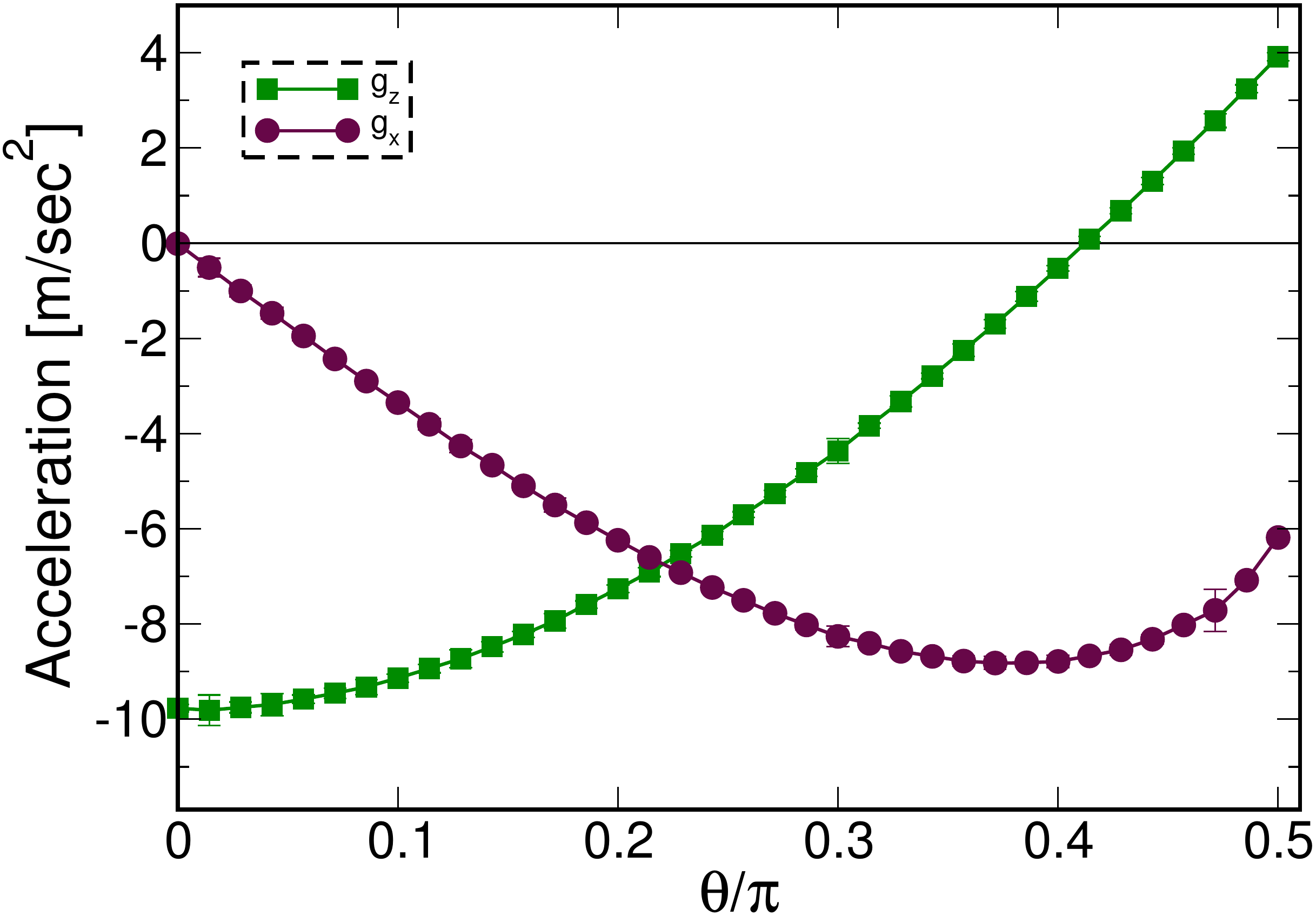}
\end{center}
\caption{\label{fig:8}Componenti dell'accelerazione gravitazionale in funzione dell'angolo $\theta$ sulla
superficie di una Terra a forma di calotta sferica.}
\end{figure}

Nella figura~\ref{fig:8} mostriamo le componenti $g_z$ e $g_x$ dell'accelerazione di gravit\`a in funzione
dell'angolo $\theta$. Notiamo come per $\theta=0$ si ha solo la componente lungo $z$, in accordo con la 
quotidiana esperienza con la forza peso. Nella regione dell'Antartide per\`o si misurerebbe una componente
tangenziale dell'accelerazione: infatti la componente normale sarebbe data da $g_x$, e $g_z\neq 0$. Inoltre
$g_z>0$ che implica che essa \`e rivolta verso il Polo Nord, quindi ancora una volta un osservatore
spostandosi verso l'Antartide lungo un meridiano sulla superficie di questa Terra misurerebbe una forza
di richiamo verso il Polo Nord, in netto contrasto con l'esperienza quotidiana.

Notiamo infine anche che per $\theta/\pi\approx 0.4$ troviamo che $g_z=0$ e $|g_x| \approx 8.67$ m/sec$^2$,
che sottostima il valore sperimentale   $g_{\mathrm{fen}} = 9.81$ m/sec$^2$ di circa il $12\%$;
per confronto, notiamo che la variazione di $g$ con la latitudine sulla Terra tonda \`e al pi\`u pari allo $0.5\%$.

\subsection{Moto di un grave sulla Terra a disco}
Abbiamo discusso nelle due sezioni precedenti di come i modelli di Terra a disco e sferica
predicano due comportamenti diversi dell'accelerazione di gravit\`a misurabili se ci si sposta
lungo un qualsiasi meridiano dal Polo Nord all'Antartide. 
A questo punto pu\`o essere utile far notare allo studente come il moto di un grave sulla superficie
della Terra apparirebbe a seconda della geometria scelta per il pianeta.
A questo scopo \`e sufficiente considerare solo i modelli di Terra sferica e a disco,
in quanto per la Terra a calotta valgono considerazioni qualitativamente in accordo con quelle
che faremo per la Terra a disco.

L'accelerazione di gravit\`a influenza il moto 
di un corpo per via della~(\ref{eq:33c}).  
Assumendo che la gravit\`a sia l'unica forza in gioco su un corpo di massa $m$ abbiamo
\begin{equation}
\bm a(\bm r_P) = \bm g(\bm r_P),
\label{eq:lp}  
\end{equation} 
dove $\bm g(\bm r_P)$ indica l'accelerazione di gravit\`a calcolata nel punto $\bm r_P$.
In questa sezione assumiamo che tutti gli spostamenti siano dell'ordine di grandezza del metro,
in modo da poter trascurare la dipendenza dell'accelerazione di gravit\`a dalla posizione\footnote{Si noti
dalla figura \ref{fig:4} che sia $g_x$ che $g_z$ variano sensibilmente
per lunghezze dell'ordine di $R_{\mathrm{MAX}}$ che \`e di circa $20\times 10^3$ km;
si pu\`o quindi trascurare la dipendenza delle accelerazioni dalle coordinate per moti
che si svolgono su scale di lunghezze pi\`u piccole come quelle che consideriamo qui.}. 

Consideriamo il moto di un corpo in caduta libera nel caso della Terra piatta: assumiamo quindi 
che il corpo inizialmente si trovi immobile vincolato ad un'altezza $z_0$ rispetto alla superficie. 
Rimuovendo il vincolo il corpo si muover\`a, per effetto della forza di gravit\`a,
di moto uniformemente accelerato lungo le direzioni $x$ e $z$.
Considerazioni cinematiche elementari
mostrano che nel tempo necessario per raggiungere il suolo, $z=0$, il grave si sposta lungo l'asse $x$ di una quantit\`a
pari a $\Delta x = x_0 - x$
\begin{equation}
\Delta x = z_0 \frac{g_x}{g_z}.
\label{eq:Dx}
\end{equation}
Troviamo quindi che alla caduta libera di un grave corrisponderebbe anche uno spostamento
in direzione longitudinale, parallelo alla superficie della Terra piatta, proporzionale al rapporto
$g_x/g_z$. Notiamo che $\Delta x>0$ per cui l'effetto della caduta libera sarebbe quello di richiamare
il grave verso il Polo Nord.

\begin{figure}[t!]
\begin{center}
\includegraphics[scale=0.35]{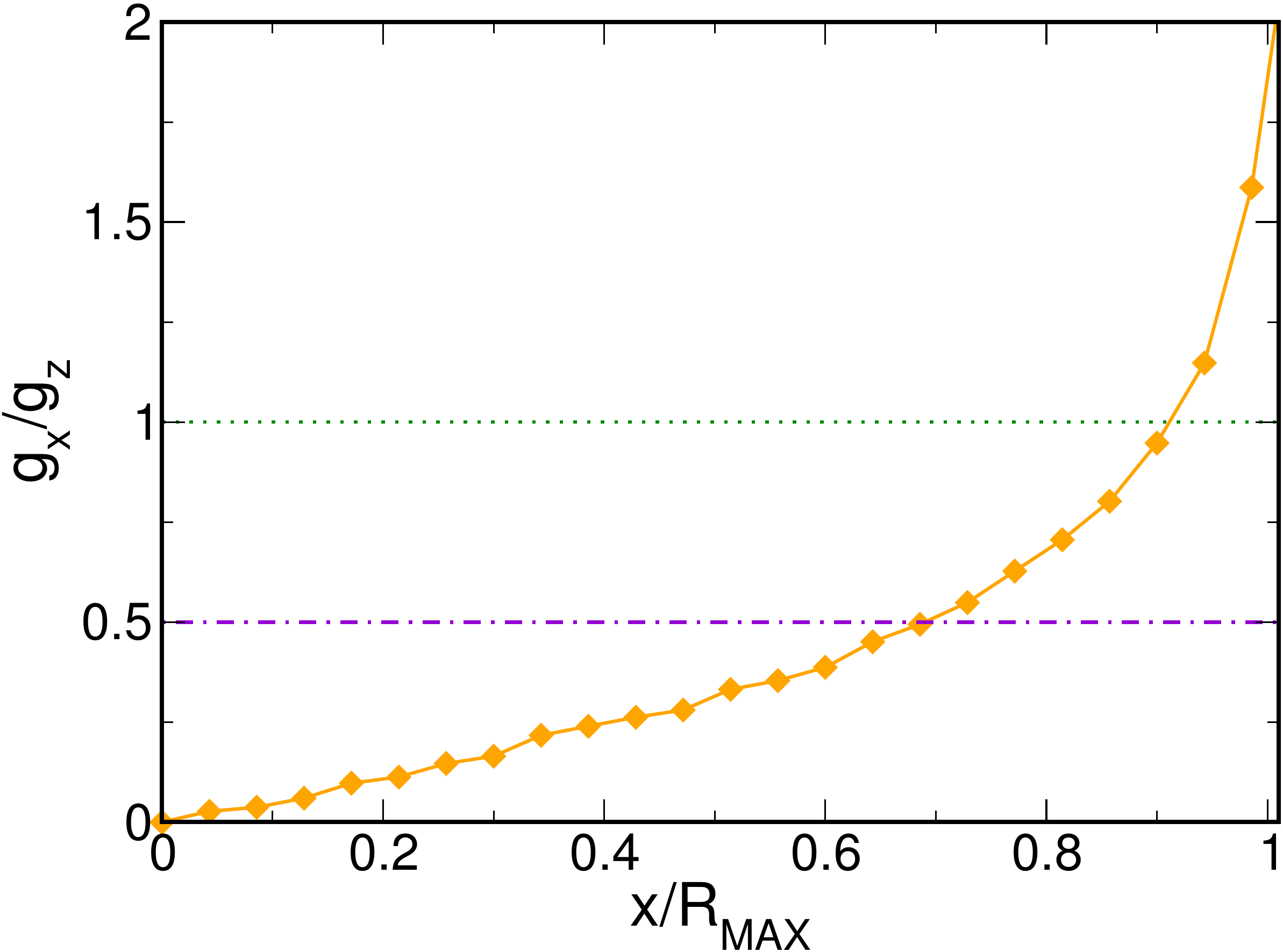}
\end{center}
\caption{\label{fig:6}Rapporto fra $g_x$ e $g_z$ nel modello Terra piatta, in funzione della distanza
misurata lungo un meridiano $x$ espressa in unit\`a del raggio della Terra $R_\mathrm{MAX}$.
Per aiutare l'occhio riportiamo anche le rette $y=1$ (linea verde tratteggiata) e 
$y=0.5$ (linea tratto-punto).}
\end{figure}

Nella figura~\ref{fig:6} mostriamo il rapporto $g_x/g_z$ in funzione di $x/R_{\mathrm{MAX}}$, che per 
la~(\ref{eq:Dx}) corrisponde allo spostamento longitudinale, espresso in unit\`a dell'altezza iniziale $z_0$,
che si misurerebbe sulla superficie della Terra 
piatta lasciando in caduta libera un grave dall'altezza $z_0$; ad esempio, vicini al bordo dove $g_x/g_z\approx 1$
si avrebbe $\Delta x\approx z_0$: lasciando cadere un grave da un metro di altezza, si dovrebbe misurare uno spostamento
longitudinale circa uguale ad un metro.
Usando i risultati riportati in figura~\ref{fig:6} si pu\`o essere pi\`u quantitativi:
fissando la quota iniziale ad un metro per semplicit\`a, troviamo che nella zona equatoriale
dove $x/R_\mathrm{MAX} \approx 0.5$ si avrebbe $\Delta x\approx 0.3$ metri: questo vuol dire che
lasciando un grave in caduta libera dalla quota di un metro in Indonesia, Repubblica Democratica del Congo, 
Nord del Brasile, Gabon, Ecuador e cos\`i via, si dovrebbe misurare uno spostamento orizzontale di circa 30 
centimetri in direzione del Polo Nord. Analogamente,  per $x/R_\mathrm{MAX} \approx 0.7$
che corrisponde alla zona del Sud di Nuova Zelanda, Argentina e Cile, si misurerebbe
uno spostamento $\Delta x\approx 0.5$ metri per ogni metro di quota. 
Il fatto che tale spostamento longitudinale non \`e mai stato 
osservato in nessun moto di caduta libera in nessuna parte del pianeta smentisce il modello 
di Terra  a disco.

\section{Conclusioni}
In questo articolo abbiamo presentato una lezione sulla gravit\`a Newtoniana che affronta il problema
delle conseguenze misurabili di una forma non sferica della Terra. Lo scopo principale della lezione
\`e mostrare allo studente come lavora il metodo scientifico, studiando un problema formalmente semplice
ma allo stesso tempo efficace in quanto basato sulla individuazione di 
grandezze fisiche misurabili sperimentalmente,
enfatizzando come la misura (o nel caso in esame, anche la semplice esperienza quotidiana) permetta
di smentire o di confermare una ipotesi. 

Nello specifico, affrontando il problema del calcolo del campo
gravitazionale che sarebbe prodotto da una Terra a disco, e confrontando il risultato con quello
che osserviamo quotidianamente, abbiamo mostrato che l'esperienza
rende il suddetto modello incompatibile con la gravit\`a Newtoniana.
In particolare, tale modello comporterebbe 
una componente tangenziale della forza di gravit\`a orientata verso il Polo Nord,
ovvero una forza di richiamo,
che causerebbe accelerazioni di gravit\`a tangenziali tanto pi\`u importanti
quanto pi\`u ci si avvicinerebbe all'Antartide.
Questa avrebbe un effetto sul moto di caduta libera di un grave:
ad esempio, lasciando cadere dalla quota di un metro un corpo, alle latitudini corrispondenti al
Sud di Nuova Zelanda, Argentina e Cile si misurerebbe 
uno spostamento orizzontale nella direzione del Polo Nord di circa cinquanta
centimetri, in netto contrasto con l'esperienza quotidiana.
 
Riteniamo che questa lezione possa essere presentata non solo durante la spiegazione della gravit\`a Newtoniana, 
ma anche in una lezione introduttiva sul metodo scientifico in generale, in quanto illustra come 
questo permetta di affrontare e risolvere un problema concreto attraverso la {\it formulazione matematica},
la {\it previsione quantitativa di grandezze fisiche} e la 
{\it verifica sperimentale delle predizioni
teoriche}. 
 
Nell'esposizione del problema e della sua soluzione non abbiamo posto l'accento sul tipo di 
algoritmo necessario per il calcolo del campo gravitazionale, sebbene questo sia basato su tecniche
standard di integrazione numerica e non presenterebbe alcuna difficilt\`a ad essere trattato
in un corso di laboratorio o di calcolo numerico.
Riteniamo infatti che nella presentazione in classe dell'argomento qui descritto, vada messo in evidenza
non tanto il calcolo quanto la capacit\`a di determinare quali grandezze fisiche 
\`e interessante studiare nel problema, studiare il risultato del calcolo e confrontarlo
con i dati sperimentali. La nostra opinione \`e che la presentazione in classe di questo argomento
non solo esercita lo spirito critico di uno studente, ma lo mette anche di fronte ad un problema 
concreto e soprattutto alla sua soluzione utilizzando il modus operandi che \`e tipico di uno scienziato.

\section*{Ringraziamenti}
Gli autori ringraziano il Dr. Giuseppe Cin\`a per le numerose discussioni sull'argomento qui presentato.
Questo lavoro \`e supportato 
dalla National Natural Science Foundation of China (grant 11575190 e 11475110) 
e dalla borsa President's International Fellowship Initiative  della Chinese Academy of Sciences 
(grant 2015PM008).

\end{document}